\documentstyle[12pt]{article} 
\begin{document} 
\begin{center}
{\bf {\huge  Three-block p-branes in various dimensions}}
\end{center}
$$~$$
\renewcommand{\theequation}{\thesection.\arabic{equation}}

\begin{center}
{\it {\bf {\Large Anastasia Volovich
}}}\\ $~$\\
{\it Department of Physics,$~$ Moscow State 
 University}\\
{\it Vorobjevi gori, Moscow, 119899}\\
{\it and}
  \\ 
 {\it Mathematical College, Independent 
  University of Moscow,}\\
 {\it 121108, PO box 68, Moscow,} {\it Russia }\footnote{e-mail 
 address:  nastya@arevol.mian.su} \end{center} \vspace{2cm} 
\begin{abstract}
It is shown that a Lagrangian, describing the interaction of the
gravitation field with the dilaton  and the antisymmetric tensor
 in arbitrary dimension 
spacetime,
admits an isotropic p-brane solution consisting of three blocks. Relations
with known  
p-brane solutions are discussed. In particular, in ten-dimensional
spacetime the three-block p-brane solution is reduced to
the known solution, which recently has been used in the 
D-brane derivation of the black hole entropy.
 
\end{abstract}

\newpage
\section{Introduction} 
\setcounter{equation}{0}

Recently, isotropic p-brane solutions in supergravity and string theory have
been extensively studied, see, for example \cite{dubholar}-\cite{berg}.
Investigation of these solutions is important for the superstring
duality \cite{tous}-\cite{schM} and for 
recent D-brane derivation 
of the black hole entropy
\cite{SV}-\cite{Hor}.
One believes that a fundamental eleven-dimensional M-theory or 
twelve-dimensional F-theory exist and 
that five known ten-dimensional superstring theories and eleven-dimensional
supergravity are low energy limits of these theories
\cite{duff1}, \cite{guven}, \cite{tous}-\cite{schM}. 
The study of p-brane solutions in an arbitrary dimension spacetime 
might   gain better understanding of the structure of superstring theory.
 
We  shall consider the following action \cite{dubholar}, \cite{duff2}:
 
\begin{equation}
 I=\frac{1}{2\kappa ^{2}}\int d^{D}x\sqrt{-g}
 (R-\frac{1}{2}(\nabla  \phi) ^{2}-
 \frac{1}{2(q+1)!}e^{-\alpha \phi}F^{2}_{q+1}),
      \label{11}
 \end{equation}
where $F_{q+1}$ is the $q+1$ differential form,
 $F_{q+1}=d{\cal A}_{q}$ and $\phi$ is the dilaton.
All ten-dimensional supergravity theories contain the terms from
 (\ref{11}) if $ D=10$, $q=2$, $\alpha =1$.
 
A class of $p$-brane solutions of (\ref{11}) for arbitrary $D$  
with the metric
 \begin{equation}
 ds^{2}=H^{k}[H^{-N}
  \eta_{\mu \nu} dy^{\mu} dy^{\nu}+
 dx^{\alpha}dx^{\alpha}], 
  \label{12}
 \end{equation}
was found in \cite{dubholar},\cite{Lu}-\cite{gibbons}.
Here parameters $k$ and $N$ are functions of the 
parameters $\alpha$, $D$ and $q$ in the action (\ref{11}).
 The solution (\ref{11}) consists of two blocks. 
The first block consists of variables $y$ and the other of variables $x$, s=
o
 we shall 
call it the two-block p-brane . This solution depends on one harmonic 
function $H(x)$.

The aim of this note is to present a solution of (\ref{11})
with the metric of the form
 \begin{equation}
 ds^{2}=H_{1}^{\frac{2}{D-2}}H_{2}^{\frac{2(D-q-2)}{q(D-2)}}
 [(H_{1}H_{2})^{-\frac{2}{q}} \eta_{\mu \nu} dy^{\mu} dy^{\nu}+
 H_{2}^{-\frac{2}{q}}dz^{m}dz^{m}+
 dx^{\alpha}dx^{\alpha}], 
  \label{13}
 \end{equation}
here $\mu$, $\nu$=0,1..., q-1; $\eta_{\mu\nu}$
 is a flat Minkowski metric; $m,n$=$1,2,...,D-2q-2$ and 
 $\alpha,\beta$ =$1,...,q+2$.
 
 The solution is valid for
 \begin{equation}
 \alpha=\pm q\sqrt{\frac{2}{D-2}}. 
  \label{14}
 \end{equation}
 This value corresponds to the supersymmetry \cite{dubholar}
  of the action (\ref{11}).
It is interesting that we get this value of $\alpha$
 not from supersymmetry, but as a solution of a system of
 algebraic equations (see Appendix B). 
 
 Here $H_{1}$ and $H_{2}$ are two arbitrary harmonic functions
 of variables $x^{\alpha}$
 \begin{equation}
  \Delta H_{1}=0,~~~~~\Delta H_{2}=0.
  \label{15}
 \end{equation}
  The dilaton field is
 \begin{equation}
  \phi=\sqrt{\frac{2}{D-2}}\ln(\frac{H_{2}}{H_{1}}).
  \label{16}
 \end{equation}
 
Non-vanishing components of the differential form are given by
 \begin{equation}
  {\cal A}_{\mu_{1}...\mu_{q}}=\sqrt{\frac{2}{q}}
  {\epsilon_{\mu_{1}...\mu_{q}}}H_{1}^{-1}
  \label{17}
 \end{equation}
 \begin{equation}
   F ^{\alpha_{1}...\alpha _{q+1}}=\sqrt{\frac{2}{q}}
   H_{1}^{-\frac{2}{D-2}}H_{2}^{-\frac{2(D-q-2)}{q(D-2)}}
   \epsilon ^{\alpha _{1}...\alpha_{q+1}\beta}
   \partial _{\beta} H_{1}^{-1}. 
  \label{18}
 \end{equation}
 Here $\epsilon _{123..q}=1$, $\epsilon ^{123..q+2}=1$

 The solution (\ref{13}) consists of three blocks, the first block 
 consists of variables $y$, another of variables $z$
 and the other of variables $x$. We shall call it
 the three-block p-branes solution.

     The solution (\ref{13}) depends on two harmonic functions.
     We shall show that for q=2 there is a three-block solution depending
     on three harmonic functions $H_{1}(x), H_{2}(x), K(x)$:
     $$
     ds^{2}=H_{1}^{-\frac{D-4}{D-2}}H_{2}^{-\frac{2}{D-2}}[-dy_{0}^{2}+
     dy_{1}^{2}+K(dy_{0}-dy_{1})^{2}]$$

 \begin{equation}
   + H_{1}^{-\frac{2}{D-2}}H_{2}^{-\frac{2}{D-2}}dz^{m}dz^{m}+
     H_{1}^{\frac{2}{D-2}}H_{2}^{\frac{D-4}{D-2}}dx^{\alpha}dx^{\alpha}
               \label{18a}
\end{equation}    
     This solution in $D=10$ dimension spacetime \cite{TS} has been considered
in recent derivation of the black hole entropy \cite {SV}-\cite{Hor} 
in the D-brane approach \cite{pol1}-\cite{pol2}.

     The paper is organized as follows.
     In Section 2 the three-block solution is derived.
      In Section  3  particular examples of the solution (\ref{13}) 
and their relations with known 
solutions are discussed. In Appendices A,B and C
 technical details are collected. 

\section{Solution of  the Einstein equations} 
 \setcounter{equation}{0}

      The Einstein equations  for the action (\ref{11}) read

\begin{equation}
  R_{MN}-\frac{1}{2}g_{MN}R=T_{MN},
  \label{19}
\end{equation}
where the energy-momentum tensor is
$$   
 T_{MN}= \frac{1}{2}
  (\partial _{M}\phi \partial _{N}\phi -
  \frac{1}{2}g_{MN} (\partial \phi)^{2}) +
$$

\begin{equation}
  \frac{1}{2q!}e^{-\alpha \phi}(F_{MM_{1}...M_{q}}F_{N}^{M_{1}...M_{q}}-
 \frac{1}{2(q+1)}g_{MN}F^{2}).\label{110}
\end{equation}

The equation of motion for the antisymmetric field is
\begin{equation}
\partial _{M}(\sqrt{-g}e^{-\alpha \phi}F^{MM_{1}...M_{q}})=0, 
                                  \label{111}
\end{equation}
where $F$ is a subject of the Bianchi identity
\begin{equation}
 \epsilon ^{M_{1}...M_{q+2}}\partial_{M_{1}}F_{M_{2}...M_{q+2}}=0.
 \label{112}
\end{equation}

The equation of motion for the dilaton is
\begin{equation}
\partial _{M}(\sqrt{-g}g^{MN}\partial _{N }\phi ) 
+ 
\frac{\alpha }{2(q+1)!}\sqrt{-g}e^{-\alpha \phi}F^{2}
=0.
 \label{113}
\end{equation}

We shall solve equations (\ref{19})-(\ref{113})  using 
the following Ansatz for the metric 

\begin{equation}
ds^{2}=e^{2A(x)}\eta_{\mu \nu} dy^{\mu} dy^{\nu}+
e^{2F(x)}\delta_{nm}dz^{n}dz^{m}+
e^{2B(x)}\delta_{\alpha\beta}dx^{\alpha}dx^{\beta}, 
 \label{114}
\end{equation}
where $\mu$, $\nu$ = 0,...,q-1, $\eta_{\mu\nu}$
is a flat Minkowski metric, $m,n$=$1,2,...,D-2q-2$ and 
 $\alpha,\beta$ =$1,...,q+2$. Here $A$, $B$
 and $C$ are functions on $x$; $\delta_{nm}$ and $\delta_{\alpha\beta}$
are Kronecker symbols. 

Non-vanishing components of the  differential forms are

\begin{equation}
{\cal A}_{\mu_{1}...\mu_{q}}=h{\epsilon_{\mu_{1}...\mu_{q}}}e^{C(x)},
 \label{115}
\end{equation}
\begin{equation}
 F ^{\alpha_{1}...\alpha _{q+1}}=\frac{1}{\sqrt{-g}}\gamma e^{\alpha 
 \phi}\epsilon ^{\alpha _{1}...\alpha_{q+1}\beta}
 \partial _{\beta} e^{\chi} , 
 \label{116}
\end{equation}
where $h,\gamma$ are constants.

 By applying the above Ansatz and using the components of the Ricci tensor 
from Appendix A, one can reduce the  $(\mu\nu)$-components of  
 (\ref{19}) to the equation 

\begin{eqnarray}
 (q-1)\Delta A+
 (q+1)\Delta B +r\Delta F  \nonumber \\ 
 +\frac{q(q-1)}{2}(\partial A)^{2} + \frac{r(r+1)}{2}(\partial F)^{2}+
 \frac {q(q+1)}{2}(\partial B)^{2} \nonumber \\ 
+q(q-1)(\partial A\partial B) + r(q-1)(\partial A\partial F) + 
rq(\partial B\partial F) = \nonumber \\ 
 -\frac{1}{4}(\partial 
 \phi)^{2} -\frac{\gamma ^{2}}{4}(\partial\chi)^{2}e^{2qB+\alpha\phi+2\chi} 
 -\frac{{h}^{2}}{4}(\partial C)^{2}e^{-\alpha\phi-2qA+2C}, 
                                     \label{117}
\end{eqnarray}

$(nm)$-components of (\ref{19}) to the following equation:

$$q\Delta A+(q+1)\Delta B+(r-1)\Delta F$$
$$+\frac{q(q+1)}{2}(\partial A)^{2}+
\frac{q(q+1)}{2}(\partial B)^{2}+\frac{r(r-1)}{2}(\partial F)^{2}$$

$$+q^{2}(\partial A\partial B)+q(r-1)(\partial A\partial F)+
q(r-1)(\partial B\partial F)=$$
\begin{equation}
 -\frac{1}{4}(\partial \phi)^{2}+\frac{h^{2}}{4}(\partial 
 C)^{2}e^{-\alpha\phi-2qA+2C}-
 \frac{\gamma^{2}}{4}(\partial \chi)^{2}e^{2qB+\alpha\phi+2\chi}
 \label{118}
\end{equation}

and $(\alpha\beta)$-components to the equation:

$$ -q\partial_{\alpha}\partial_{\beta} A-
q\partial_{\alpha}\partial_{\beta} B-
r\partial_{\alpha}\partial_{\beta} F $$
$$-q\partial_{\alpha} A\partial_{\beta} A + q\partial_{\alpha} 
B\partial_{\beta} B - r\partial_{\alpha} F\partial_{\beta} F + 
q(\partial_{\alpha} A\partial_{\beta} B +\partial_{\alpha} 
B\partial_{\beta} A)$$
$$+r(\partial_{\alpha} B\partial_{\beta} F + \partial_{\alpha} 
F\partial_{\beta} B) + \delta_{\alpha\beta}[q\Delta A+q\Delta B+
r\Delta F$$
$$+\frac{q(q+1)}{2}(\partial A)^{2} + \frac{r(r+1)}{2}(\partial F)^{2}  $$
$$ +\frac{q(q-1)}{2}(\partial B)^{2} +q(q-1)(\partial A\partial B) + 
r(q-1)(\partial F\partial B )+ qr(\partial A\partial F)] =$$

$$=\frac{1}{2}\partial_{\alpha} \phi\partial_{\beta} \phi - 
\frac{1}{4}\delta_{\alpha\beta}(\partial\phi)^{2} - 
\frac{h^{2}}{2}e^{-\alpha\phi-2qA+2C}[\partial_{\alpha} C\partial_{\beta} C
-\frac{\delta_{\alpha\beta}}{2}(\partial C)^{2}]$$

\begin{equation}
-\frac{\gamma^{2}}{2} e^{2qB+\alpha\phi+
2\chi}[\partial_{\alpha} \chi\partial_{\beta} \chi 
 -\frac{\gamma{\delta_{\alpha\beta}}}{2}(\partial \chi)^{2}],
 \label{119}
\end{equation}
where we use notations 
$( \partial A\partial B)=\partial_{\alpha} A\partial_{\alpha} F$
and $r=D-2q-2$.

The equations of motion (\ref{111}) for a part of components of the
 antisymmetric field are identicaly satisfied and for the other part 
  they are reduced to a simple equation:

\begin{equation}
\partial _{\alpha}(e^{-\alpha \phi -2qA +C}\partial _{\alpha}C)=0.
                  \label{120}
\end{equation}

For $\alpha$-components of the antisymmetric field we 
 also have the Bianchi identity:

\begin{equation}
\partial_{\alpha}(e^{\alpha\phi + 2Bq + \chi }\partial_{\alpha}\chi)=0. 
 \label{121}
\end{equation}

The equation of motion for the dilaton has the form

$$\partial _{\alpha}(e^{qA + qB + Fr}\partial _{\alpha}\phi )
+\frac{\alpha \gamma ^{2}}{2}e^{\alpha \phi +Aq + 3qB +rF +2\chi}
(\partial _{\alpha}\chi)^{2} $$

\begin{equation}
-\frac{\alpha h^{2}}{2}e^{-\alpha \phi -qA + qB +rF
+2C}(\partial _{\alpha} C)^{2}=0.
  \label{122}
\end{equation}

We have to solve the system of equations (\ref{117})-(\ref{122})
for unknown functions
$ A,B,F,C,\chi,\phi$. We shall express $ A,B,F$ and  $\phi $
in terms of two functions $C$ and $\chi$. 
In order to kill exponents in (\ref{117})-(\ref{122}) 
we impose the following relations:
\begin{equation}
qA + rF +qB =0,
 \label{123}
\end{equation}
\begin{equation}
\alpha \phi + 2\chi + 2qB =0,
 \label{124}
\end{equation}
\begin{equation}
 2C - 2qA -\alpha\phi = 0.
 \label{125}
\end{equation}
Under these conditions equations (\ref{120}),(\ref{121}) and (\ref{122})
will have the following forms, respectively,

\begin{equation}
\partial _{\alpha}(e^{-C}\partial _{\alpha}C)=0,~~~~~
\partial _{\alpha}(e^{-\chi}\partial _{\alpha}\chi)=0,
                                   \label{126}
\end{equation}
\begin{equation}
\Delta \phi +\frac{\alpha \gamma ^{2}}{2}(\partial _{\alpha}  \chi )^{2}-
\frac{\alpha  h^{2}}{2}(\partial _{\alpha} C )^{2}=0.
                  \label{127}
\end{equation}

One rewrites (\ref{126}) as
\begin{equation}
 \Delta C =(\partial C)^{2},~~~~~
 \Delta \chi =(\partial \chi)^{2}.                  \label{128}
\end{equation}
Therefore (\ref{127}) will have the form
\begin{equation}
\Delta \phi +\frac{\alpha \gamma ^{2}}{2}\Delta \chi - 
\frac{\alpha  h^{2}}{2}\Delta C =0.
                      \label{129}
\end{equation}

From (\ref{129}) it is natural to set
\begin{equation}
\phi =\phi_{1}C + \phi_{2}\chi, 
 \label{130}
\end{equation}
where 
\begin{equation}
 \phi_{1}=\frac{\alpha h^{2}}{2},~~~
 \phi_{2}=-\frac{\alpha\gamma^{2}}{2}.
 \label{131}
\end{equation}

From equations (\ref{123}), (\ref{124}) and (\ref{125})  it follows that 
$A$, $B$ and $F$ can 
be presented as 
linear combinations of functions $C$ and $\chi$:

\begin{equation}
 A=a_{1}C + a_{2}\chi ,
 \label{132}
\end{equation}
\begin{equation}
 B=b_{1}C + b_{2}\chi ,
 \label{133}
\end{equation}
\begin{equation}
 F=f_{1}C + f_{2}\chi ,
 \label{134}
\end{equation}
where

\begin{equation}
 a_{1}=\frac{4-\alpha^{2}h^{2}}{4q}, ~~~
 a_{2} = \frac{\alpha^{2}\gamma^{2}}{4q},
 \label{135}
\end{equation}

\begin{equation}
 b_{1} = -\frac{\alpha^{2}h^{2}}{4q},~~~
 b_{2} =\frac{\alpha^{2}\gamma^{2}-4}{4q},
  \label{136}
\end{equation}

\begin{equation}
 f_{1} = \frac{\alpha^{2}h^{2} - 2}{2r},~~~
 f_{2} = \frac{2-\alpha^{2}\gamma^{2}}{2r}.
 \label{137}
\end{equation}

Let us substitute expressions (\ref{130}),(\ref{132})-(\ref{134}) for
$\phi,A,B,F$ into (\ref{117})-(\ref{119}). We get relations containing 
bilinear forms over derivatives on $ C$ and $\chi$ .
 We assume that the
coefficients in front of these  bilinear
forms vanish. Therefore we get the system of twelve quartic
(actually quadratic) equations for three unknown parameters 
$\alpha$, $h $ and $\gamma$.
These relations are presented and solved in Appendix B.
The solution is:
\begin{equation}
 { \alpha}^{2}={q}^{2}\frac{2}{r+2q} ,
  ~~{h}^{2} ={\gamma}^{2} =\frac{2}{q},
 \label{138}
\end{equation}
\begin{equation}
 a_{1}=\frac{q+r}{q(r+2q)},~~~~~a_{2}=\frac{1}{r+2q},
 \label{139}
\end{equation}
\begin{equation}
 b_{1}=-\frac{1}{r+2q},~~~~~b_{2}=-\frac{q+r}{q(r+2q)},
  \label{140}
\end{equation}
\begin{equation}
 f_{1}=-\frac{1}{r+2q},~~~~~f_{2}=\frac{1}{r+2q},
 \label{141}
\end{equation}

Finally, we get 8 different solutions which differ in terms 
of signs $\alpha, h, \gamma$.
They are 

$$h=\pm \sqrt{\frac{2}{q}},~~~\gamma=\pm 
\sqrt{\frac{2}{q}},~~~\alpha= \pm q\sqrt{\frac{2}{r+2q}}.$$

From (\ref{131}) we get that 
 
$${\phi}_{1}=\frac{\alpha}{q}~~~~ 
 \mbox{and}~~~~{\phi}_{2}=-\frac{\alpha}{q},$$
so
\begin{equation}
 \phi_{1}=\pm \sqrt{\frac{2}{r+2q}},~~~~~\phi_{2}=\mp \sqrt{\frac{2}{r+2q}}.
                                    \label{142}
\end{equation}

Unknown functions in the metric (\ref{114}) were presented in terms of
functions $C$ and $\chi$, which satisfy the equations (\ref{128}). 
Let us express $C$ and $\chi$ in terms of $H_{1}$ and $H_{2}$ by using
the formulas:
\begin{equation}
 e^{-C}=H_{1},~~~~~~~e^{-\chi}=H_{2}. 
  \label{143}
\end{equation}
Then from (\ref{128})
 one gets that $H_{1}$ and $H_{2}$ are harmonic functions
 \begin{equation}
  \Delta H_{1}=0,~~~~~~~\Delta H_{2}=0.
  \label{144}
 \end{equation}
 By using expresions for $a_{1},a_{2},b_{1},b_{2},f_{1},f_{2},\phi_{1},
 \phi_{2}$
 in terms  of $\alpha,h,\gamma$ and $C,\chi$ in terms of $H_{1}$ and $H_{2}$
 one gets: 
 
 \begin{equation}
  A=-\frac{D-q-2}{q(D-2)}\ln H_{1}-\frac{1}{D-2}\ln H_{2},
  \label{145}
 \end{equation}
 
 \begin{equation}
  B=\frac{1}{D-2}\ln H_{1}+\frac{D-q-2}{q(D-2)}\ln H_{2},
  \label{146}
 \end{equation}
 
  \begin{equation}
  F=\frac{1}{D-2}\ln H_{1}-\frac{1}{D-2}\ln H_{2},
  \label{147}
 \end{equation}
 
  \begin{equation}
  \phi=\mp \sqrt{\frac{2}{D-2}}\ln H_{1} \pm \sqrt{\frac{2}{D-2}}\ln H_{2}.
  \label{148}
 \end{equation}
  
 Finally, by using (\ref{145})-(\ref{147}) we get the expression for 
 the metric (\ref{114})
 
 $$ ds^{2}=H_{1}^{\frac{-2(D-q-2)}{q(D-2)}}H_{2}^{\frac{-2}{D-2}} 
 \eta_{\mu \nu} dy^{\mu} dy^{\nu}+
$$

 \begin{equation}
 H_{1}^{\frac{2}{D-2}}H_{2}^{-\frac{2}{D-2}}dz^{m}dz^{m}+
  H_{1}^{\frac{2}{D-2}}H_{2}^{\frac{2(D-q-2)}{q(D-2)}}
  dx^{\alpha}dx^{\alpha}, 
  \label{149}
 \end{equation}
that is equivalent to (\ref{13}) presented in Introduction.

\section{Particular cases}
\setcounter{equation}{0}

In this section different particular cases of the solution (\ref{149})
 are discussed.
\subsection *{Boosted solution} 
For $q=2$ equations (\ref{19}) admit  a solution depending
on three harmonic functions.
Let us use the following ansatz for the metric 

$$ds^{2}=e^{2A(x)}[-dy^{2}_{0}+dy^{2}_{1}+K(dy_{0}-dy_{1})^{2}]$$

\begin{equation}
+e^{2F(x)}dz^{m}dz^{m}
+e^{2B(x)}dx^{\alpha}dx^{\alpha},  
 \label{150}
\end{equation}
where $K$ is a function of variables $x$. We shall use ansatz 
(\ref{116}), (\ref{117}) for 
the antisymmetric field and coordinates 
$$v=y_{1}+y_{0},~~~~~u=y_{0}-y_{1}.$$

For this ansatz the form of equations of motion (\ref{121}) and (\ref{122})
as well as the form of  $uv$, $vv$ and also  $mn$ and $\mu\nu$-components of 
the Einstein equations does not change. The form for $uu$-component
changes.  
  For details see Appendix C.
 For example, for  $uv$-components of the Einstein 
 equations we have
\begin{eqnarray}
-\Delta A+\Delta B  + 
(\partial A)^{2}+
\frac{r(r+1)}{2}(\partial F)^{2}+3(\partial B)^{2} \nonumber \\ 
+2(\partial A\partial B)+
r(\partial A\partial F) +
2r(\partial F\partial B) = 
 -\frac{1}{4}(\partial 
 \phi)^{2} -\frac{\gamma ^{2}}{4}(\partial C)^{2}
 -\frac{\gamma ^{2}}{4}(\partial \chi)^{2}.
 \label{151}
\end{eqnarray}
Here we use the relation
\begin{equation}
 2A+rF+2B=0.
 \label{152}
\end{equation}
 The Einstein equations for  $uu$-component  has the form
\begin{eqnarray}
 -\frac{1}{2}\Delta K+K[-\Delta A+\Delta B  + 
(\partial A)^{2}+
\frac{r(r+1)}{2}(\partial F)^{2}+3(\partial B)^{2}\nonumber \\ 
+2(\partial A\partial B)+
r(\partial A\partial F) +
2r(\partial F\partial B) ]
-\partial K [2\partial A+r\partial F+2\partial B]=
                                                 \nonumber \\ 
 -K[\frac{1}{4}(\partial \phi)^{2}+ 
\frac{\gamma ^{2}}{4}(\partial C)^{2}+
\frac{\gamma ^{2}}{4}(\partial \chi)^{2}].
                          \label{153}
\end{eqnarray}

We see that in equation (\ref{153}) the terms containing $K$ without 
derivatives are canceled due to equation (\ref{151}). Terms with the 
first order derivative are canceled  due to  relation (\ref{152}). 
Therefore, if we assume that 
\begin{equation}
\Delta K=0, 
 \label{154}
\end{equation} 
we get that the metric

$$     ds^{2}=H_{1}^{-\frac{D-4}{D-2}}H_{2}^{-\frac{2}{D-2}}[-dy_{0}^{2}+
     dy_{1}^{2}+K(dy_{0}-dy_{1})^{2}$$

 \begin{equation}
  +H_{1}^{-\frac{2}{D-2}}H_{2}^{-\frac{2}{D-2}}dz^{m}dz^{m}+
     H_{1}^{\frac{2}{D-2}}H_{2}^{\frac{D-4}{D-2}}dx^{\alpha}dx^{\alpha}]
      \label{155}
     \end{equation}
    solves the theory.

\subsection *{Magnetic charge}
Let us take $H_{1}=1,~H_{2}=H$. This ansatz corresponds to
zero electic field (see (\ref{115})) and non-zero magnetic charge. 
In this case the three-block metric reduces to two-block metric
\begin{equation}
ds^{2}=H^{-\frac{2}{D-2}}[\eta _{\mu \nu} dy^{\mu} dy^{\nu}+
dz^{m}dz^{m}]
+H^{2\frac{D-2-q}{q(D-2)}}dx^{\alpha}dx^{\alpha}. 
 \label{156}
\end{equation}
More general solution for arbitrary $\alpha$ 
has been presented in \cite{dubholar}, \cite{Lu}-\cite{gibbons},
\cite{Tseytlin}.
The metric is:
 
 \begin{equation}
  ds^{2}=H^{\sigma}[H^{-N}\eta_{\mu\nu}dy^{\mu}dy^{\nu}+
  dx^{\alpha}dx^{\alpha}].
  \label{157}
 \end{equation}
 $$N=\frac{4}{\Delta},~~~~~\sigma=\frac{4(D-q-2)}{\Delta(D-2)},$$
 
 $$~~~~~\Delta\equiv a^{2}+\frac{2q(D-q-2)}{D-2}.$$
 
 Here $\mu,\nu$=0,1...,D-q-3;~~$\eta_{\mu\nu}$ is a flat
 Minkovski metric and $\alpha$=$1,2...,q+2$.
It depends on one harmonic function $H(x)$.

\subsection *{Electric charge}

Let us take $H_{1}=H,~H_{2}=1$. This ansatz corresponds to
zero magnetic field  and non-zero electic charge. 
In this case the three-block metric reduces to two-block metric
\begin{equation}
ds^{2}=H^{-2\frac{D-2-q}{q(D-2)}}\eta _{\mu \nu} dy^{\mu} dy^{\nu}+
H^{\frac{2}{D-2}}[dz^{m}dz^{m}+dx^{\alpha}dx^{\alpha}],  
 \label{158}
\end{equation}
More general solution for arbitrary $\alpha$ has been presented in 
\cite{dubholar,AVV}.

\subsection *{Non-dilaton solution}

Let us take $H_{1}=H_{2}=H$. This ansatz corresponds to
equal 
electic and  magnetic chargres. 
In this case the three-block metric reduces to a two-block metric
plus flat Euclidean metric
\begin{equation}
ds^{2}=H^{-\frac{2}{q}}\eta _{\mu \nu} dy^{\mu} dy^{\nu}+dz^{m}dz^{m}
+H^{\frac{2}{q}}dx^{\alpha}dx^{\alpha}.  
 \label{159}
\end{equation}

\subsection *{Self-dual solution}
In the particular case of $D=2q+2$ magnetic and electric fields 
fill all spacetime, and the three-block metric reduces 
to the two-block metric 
\begin{equation}
ds^{2}=(H_{1}H_{2})^{-\frac{1}{q}}\eta _{\mu \nu} dy^{\mu} dy^{\nu}+
(H_{1}H_{2})^{\frac{1}{q}}dx^{\alpha}dx^{\alpha},  
 \label{160}
\end{equation}
For this particular case $\alpha=\pm \sqrt{q}$.
\subsection *{ D=4 and q=1}
 If D=4 and q=1 then 
  \begin{equation}
   ds^{2}=-(H_{1}H_{2})^{-1}dy_{0}^{2}+
   H_{1}H_{2}(dx_{1}^{2}+dx_{2}^{2}+dx_{3}^{2})
   \label{161}
  \end{equation}
This metric was obtained in \cite{kallosh} for an action with two 
antisymmetric fields.

\subsection *{D=10, q=2, K$\neq $0}

If we take $D=10$, $q=2$, $K\neq 0$ then we get the  solution
\cite{TS,CM,CH}
$$  ds^{2}=H_{1}^{-\frac{3}{4}}H_{2}^{-\frac{1}{4}}
  (-dy_{0}^{2}+dy_{1}^{2}+K(dy_{0}- dy_{1})^{2})
$$ 
\begin{equation}
 + H_{1}^{\frac{1}{4}}H_{2}^{-\frac{1}{4}}
  (dz_{1}^{2}+dz_{2}^{2}+dz_{3}^{2}+dz_{4}^{2})+
  H_{1}^{\frac{1}{4}}H_{2}^{\frac{3}{4}}
  (dx_{1}^{2}+dx_{2}^{2}+dx_{3}^{2}+dx_{4}^{2}).
  \label{162}
 \end{equation}


\subsection *{D=10, q=1}
Let us also mention the following solution for $D=10$, $q=1$
 
$$ ds^{2}=H_{1}^{\frac{1}{4}}H_{2}^{\frac{7}{4}}
 [-(H_{1}H_{2})^{-{2}}{dy_{0}}^{2} +$$
 
 $$H_{2}^{-2}(dz_{1}^{2}+dz_{2}^{2}+dz_{3}^{2}+dz_{4}^{2}+dz_{5}^{2}+
 dz_{6}^{2})+$$
 \begin{equation}
dx_{1}^{2}+dx_{2}^{2}+dx_{3}^{2}], 
  \label{163}
 \end{equation}

\section{Concluding Remarks}

To conclude, we have constructed the three-block solution (\ref{13})
of the Lagrangian (\ref{11}) and discussed its reduction to known
solutions. 
It is interesting that the value of the dilaton parameter $\alpha$
(\ref{14})
was obtained as a solution of the system of algebraic equations which 
follows from the 
Einstein equations for the three-block
metric (\ref{114}).  There is no such fixing for the two-block p-brane 
(\ref{12}). As it is known, this value  of $\alpha$ leads to a 
supersymmetric theory \cite{dubholar}.
It would be interesting to gain better undestanding of a mechanism 
responsible for this fixing.
 It would be also interesting to generalize the three-block solution to 
 n-block  solutions ($n\geq 4$) and to see the connection with the harmonic
 superposition of M-branes
 \cite {papadopoulos,Tseytlin,klebanov}
\section*{Acknowlegment}

I am grateful to K.S.Viswanathan for
the invitation to visit Physics Department of Simon Fraser University,
Vancouver, Canada, where this work was done. I am especially 
grateful to I.Ya.Arefeva and I.V.Volovich for suggesting this problem to me
and encouraging me to pursue this study. I am very grateful to 
K.Behrndt who pointed out a missing solution for the parameter 
$\alpha$ in the previous version of the paper. I am also grateful to 
E.Bergshoeff for helpful remarks.


\newpage

\appendix
\section *{Appendix}
\setcounter{equation}{0}

\section {Three-block metric}
Let us calculate the left hand side of the Einstein equations for the metric
\begin{equation}
ds^{2}=e^{2A(x)}\eta_{\mu \nu} dy^{\mu} dy^{\nu}+e^{2F(x)}dz^{m}dz^{m}
+e^{2B(x)}dx^{\alpha}dx^{\alpha}, 
 \label{164}
\end{equation}
$\mu$, $\nu$ =$0$,... $q-1$, $\eta_{\mu\nu}$
is a flat Minkowski metric with signature $(-,+...+)$;
 $m,n$ = $1,...r$; $r=d-q$;
$\alpha,\beta$= 
$1$,...$D-d$.
$A$, $B$
 and $C$ are functions on $x$.

 For this metric  non-vanishing  Christoffel symbols have the form:
 \begin{equation}
  \Gamma^{\mu}_{\nu\alpha}=\delta_{\mu}^{\nu}\partial _{\alpha} A,
                              \label{165}
 \end{equation}
 \begin{equation}
  \Gamma^{\alpha}_{\mu\nu}=-h_{\mu\nu}e^{2(A-B)}\partial_{\alpha }A,
  \label{166}
 \end{equation}
 \begin{equation}
  \Gamma^{\alpha}_{mn}=-\delta_{mn}e^{2(F-B)}\partial_{\alpha} F,
  \label{167}
 \end{equation}
  \begin{equation}
  \Gamma^{m}_{n\alpha}=\delta^{m}_{n}\partial_{\alpha} F,
  \label{168}
 \end{equation}
\begin{equation}
 \Gamma^{\alpha}_{\beta\gamma}=\delta^{\alpha}_{\beta}\partial_{\gamma}
 B+\delta^{\alpha}_{\gamma}\partial_{\beta} 
 B-\delta_{\beta}^{\gamma}\partial_{\alpha} B.
 \label{169}
\end{equation} 
The Riemann tensor is defined as

\begin{equation}
 R^{M}_{NKL}=\partial_{K}\Gamma^{M}_{NL}-\partial_{L}\Gamma^{M}_{NK}+
 \Gamma^{M}_{PK}\Gamma^{P}_{NL}-\Gamma^{M}_{PL}\Gamma^{P}_{NK}.
 \label{170}
\end{equation}
The Ricci tensor is
$$R_{MK}=R^{N}_{MNK}.$$

For the metric (\ref{164}) the components of the Ricci tensor are:
\begin{equation}
 R_{\mu\nu}=-h_{\mu\nu}e^{2(A-B)}[\Delta  A +q(\partial A)^{2}
  +{\tilde d}(\partial A\partial B)+r(\partial A\partial F)],
 \label{171}
\end{equation}

\begin{equation}
 R_{mn}=-\delta_{mn}e^{2(F-B)}[\Delta  F +r(\partial F)^{2}
  +q(\partial A\partial F)+{\tilde d}(\partial B\partial F)],
 \label{172}
\end{equation}

$$R_{\alpha\beta}=-q\partial_{\alpha}\partial_{\beta} A-
r\partial_{\alpha}\partial_{\beta} F -
 {\tilde d}\partial_{\alpha}\partial_{\beta} B-
 q\partial_{\alpha} A\partial_{\beta} A -r\partial_{\alpha} 
F\partial_{\beta} F
+{\tilde d}\partial_{\alpha} B\partial_{\beta} B$$    

 $$+q(\partial_{\alpha} A\partial_{\beta} B+
 \partial_{\alpha} B\partial_{\beta} A)+
 r(\partial_{\alpha} B\partial_{\beta} F +
 \partial_{\alpha} F\partial_{\beta} B)$$

\begin{equation}
 +\delta_{\alpha\beta}[-\Delta B-{\tilde d}(\partial B)^{2}-q
 (\partial A\partial B) - r(\partial F\partial B)],
 \label{173}
\end{equation}
where ${\tilde d}=D-d-2,~r=d-q$.

Scalar curvature is
$$R=e^{-2B}[ -2q\Delta A 
 -2({\tilde d}+1)\Delta B -2r\Delta F$$

$$ -q(q+1)(\partial A)^{2} 
 -{\tilde d}({\tilde d}+1)(\partial B)^{2}-r(r+1)
 (\partial F)^{2}$$
 
\begin{equation}
-2qr(\partial A\partial F)-
2q{\tilde d}(\partial A\partial B)-
 2r{\tilde d}(\partial B\partial F)].
 \label{174}
\end{equation}

 The left hand side of the Einstein equations read

$$R_{\mu\nu}-\frac{1}{2}g_{\mu\nu}R=\eta_{\mu\nu}e^{2(A-B)}[(q-1)\Delta A
+({\tilde d}+1)\Delta B + r\Delta F  $$

$$+\frac{q(q-1)}{2}(\partial A)^{2}+
\frac{r(r+1)}{2}(\partial F)^{2}+\frac{{\tilde d}({\tilde 
d}+1)}{2}(\partial B)^{2}$$

\begin{equation}
+{\tilde d}(q-1)(\partial A\partial 
B)+r(q-1)(\partial A\partial F) +r{\tilde d}(\partial F\partial B),
 \label{175}
\end{equation}

$$R_{mn}-\frac{1}{2}g_{mn}R=\delta_{mn}e^{2(F-B)}[q\Delta A+
({\tilde d}+1)\Delta B + (r-1)\Delta F $$

$$ +\frac{q(q+1)}{2}(\partial A)^{2}+
\frac{r(r-1)}{2}(\partial F)^{2}+\frac{{\tilde d}({\tilde 
d}+1)}{2}(\partial B)^{2}$$

\begin{equation}
+{\tilde d}q(\partial A\partial 
B)+q(r-1)(\partial A\partial F) +{\tilde d}(r-1)(\partial F\partial B)],
 \label{176}
\end{equation}

$$R_{\alpha \beta} -\frac{1}{2} g_{\alpha \beta} R 
=-q\partial_{\alpha}\partial_{\beta} A-
{\tilde d}\partial_{\alpha}\partial_{\beta} B-
r\partial_{\alpha}\partial_{\beta} F$$

$$-q\partial_{\alpha} A\partial_{\beta}A+
{\tilde d}\partial_{\alpha} B\partial_{\beta}B-
r\partial_{\alpha} F\partial_{\beta} F$$

$$+q(\partial_{\alpha} A\partial_{\beta} B+
\partial_{\alpha} B\partial_{\beta} A)+
r(\partial_{\alpha} B\partial_{\beta} F+
\partial_{\alpha} F\partial_{\beta} B)$$

$$+\delta_{\alpha\beta}[q\Delta A+
{\tilde d}\Delta B + r\Delta F + \frac{q(q+1)}{2}(\partial A)^{2}+
\frac{r(r+1)}{2}(\partial F)^{2}+
\frac{{\tilde d}({\tilde d}-1)}{2}(\partial B)^{2}$$

 \begin{equation}
+q({\tilde d}-1)(\partial A\partial B)+
qr(\partial A\partial F) +r({\tilde d}-1)(\partial F\partial B)].
 \label{177}
\end{equation}

\newpage

\section{System of algebraic equations}
Here we present a system of equations which follows from 
equations (\ref{117}) - (\ref{119}). This system is

$$-a_{1}+b_{1}+\frac{q(q-1)}{2}a_{1}^{2}+\frac{r(r+1)}{2}f_{1}^{2}+
\frac{q(q+1)}{2}b_{1}^{2}$$

\begin{equation}
 +q(q-1)a_{1}b_{1}+r(q-1)a_{1}f_{1}+rqf_{1}b_{1}+\frac{\phi_{1}^{2}}{4}+
 \frac{h^{2}}{4}=0;
 \label{178}
\end{equation}

$$-a_{2}+b_{2}+\frac{q(q-1)}{2}a_{2}^{2} + 
\frac{r(r+1)}{2}f_{2}^{2} + \frac{q(q+1)}{2}b_{2}^{2}$$

\begin{equation}
 +q(q-1)a_{2}b_{2}+r(q-1)a_{2}f_{2}+rqb_{2}f_{2} 
 +\frac{\phi_{2}^{2}}{4}+\frac{\gamma^{2}}{4}=0;
 \label{179}
\end{equation}

$$q(q-1)a_{1}a_{2} +r(r+1)f_{1}f_{2}+q(q+1)b_{1}b_{2}$$

\begin{equation}
 +q(q-1)(a_{1}b_{2}+a_{2}b_{1})+r(q-1)(a_{2}f_{1}+a_{1}f_{2})+rq(f_{1}b_{2}
 +f_{2}b_{1})+\frac{\phi_{1}\phi_{2}}{2}=0;
 \label{180}
\end{equation}

$$-f_{1}+b_{1}+\frac{q(q+1)}{2}a_{1}^{2}+\frac{r(r-1)}{2}f_{1}^{2}+
\frac{q(q+1)}{2}b_{1}^{2}$$

\begin{equation}
 +q^{2}a_{1}b_{1}+q(r-1)a_{1}f_{1}+q(r-1)f_{1}b_{1}+\frac{\phi_{1}^{2}}{4}
 -\frac{h^{2}}{4}=0;
 \label{181}
\end{equation}

$$-f_{2}+b_{2}+\frac{q(q+1)}{2}a_{2}^{2} + 
\frac{r(r-1)}{2}f_{2}^{2} + \frac{q(q+1)}{2}b_{2}^{2}+$$

\begin{equation}
 +q^{2}a_{2}b_{2}+q(r-1)a_{2}f_{2}+q(r-1)b_{2}f_{2} 
 +\frac{\phi_{2}^{2}}{4}+\frac{\gamma^{2}}{4}=0;
 \label{182}
\end{equation}

$$q(q+1)a_{1}a_{2} +r(r-1)f_{1}f_{2}+q(q+1)b_{1}b_{2}$$

\begin{equation}
 +q^{2}(a_{1}b_{2}+a_{2}b_{1})+q(r-1)(a_{2}f_{1}+a_{1}f_{2})
 +q(r-1)(f_{1}b_{2}
 +f_{2}b_{1})+\frac{\phi_{1}\phi_{2}}{2}=0;
 \label{183}
\end{equation}

$$-qa_{1}^{2}-rf_{1}^{2}+
qb_{1}^{2}+$$

\begin{equation}
 +2qa_{1}b_{1}+2rf_{1}b_{1}-\frac{\phi_{1}^{2}}{2}+
 \frac{h^{2}}{2}=0;
 \label{184}
\end{equation}

$$-qa_{2}^{2} -rf_{2}^{2} + qb_{2}^{2}$$

\begin{equation}
 +2qa_{2}b_{2}+2rb_{2}f_{2} 
 -\frac{\phi_{2}^{2}}{2}+\frac{\gamma^{2}}{2}=0;
 \label{185}
\end{equation}

$$-qa_{1}a_{2} -rf_{1}f_{2}+qb_{1}b_{2}$$

\begin{equation}
 +q(a_{1}b_{2}+a_{2}b_{1})+r(f_{1}b_{2}
 +f_{2}b_{1})-\frac{\phi_{1}\phi_{2}}{2}=0;
 \label{186}
\end{equation}

$$\frac{q(q+1)}{2}a_{1}^{2}+\frac{r(r+1)}{2}f_{1}^{2}+
\frac{q(q-1)}{2}b_{1}^{2}$$

\begin{equation}
 +q(q-1)a_{1}b_{1}+r(q-1)b_{1}f_{1}+rqf_{1}a_{1}+\frac{\phi_{1}^{2}}{4}-
 \frac{h^{2}}{4}=0;
 \label{187}
\end{equation}

$$\frac{q(q+1)}{2}a_{2}^{2} + 
\frac{r(r+1)}{2}f_{2}^{2} + \frac{q(q-1)}{2}b_{2}^{2}$$

\begin{equation}
 +q(q-1)a_{2}b_{2}+r(q-1)a_{2}f_{2}+rqb_{2}f_{2} 
 +\frac{\phi_{2}^{2}}{4}-\frac{\gamma^{2}}{4}=0;
 \label{188}
\end{equation}

$$q(q+1)a_{1}a_{2} +r(r+1)f_{1}f_{2}+q(q-1)b_{1}b_{2}+
q(q-1)(a_{1}b_{2}+a_{2}b_{1})
$$

\begin{equation}
+r(q-1)(b_{2}f_{1}+b_{1}f_{2})+rq(f_{1}a_{2}
 +f_{2}a_{1})+\frac{\phi_{1}\phi_{2}}{2}=0.
 \label{188a}
\end{equation}
The variables $a_{1},a_{2},b_{1},b_{2},f_{1},f_{2},\phi_{1},\phi_{2}$
are functions of $h,\alpha,\gamma$. So we have a system of twelve 
quadratic equations for
three unknown parametres $h,\alpha,\gamma$. We solved this system
of equations on MAPLE V. It occurs that the system has solutions
only for special value of $\alpha$, 
\begin{equation}
 \alpha=\pm q\sqrt{\frac{2}{D-2}}
 \label{189}
\end{equation}

The solution is:
\begin{equation}
 h^{2}={\gamma}^{2}=\frac{2}{q},  
 \label{190}
\end{equation}
\begin{equation}
 a_{1}=\frac{q+r}{q(r+2q)},~~~~~a_{2}=\frac{1}{r+2q},
 \label{191}
\end{equation}

\begin{equation}
 b_{1}=-\frac{1}{r+2q},~~~~~b_{2}=-\frac{q+r}{q(r+2q)}
  \label{192}
\end{equation}

\begin{equation}
 f_{1}=-\frac{1}{r+2q},~~~~~f_{2}=\frac{1}{r+2q},
 \label{193}
\end{equation}

\begin{equation}
{\phi_{1}}^{2}=\frac{2}{r+2q},~~~~~{\phi_{2}}^{2}=\frac{2}{r+2q}.
 \label{194}
\end{equation}

\newpage
\section{Boosted three-block metric}
It is convinient to rewrite the metric (\ref{150}) in the light-cone
coordinates. 
$v=y_{0}+y_{1},~u=y_{0}-y_{1},$
\begin{equation}
ds^{2}=e^{2A(x)}[-dudv+K(x)du^{2}]+
e^{2F(x)}dz^{m}dz^{m}
+e^{2B(x)}dx^{\alpha}dx^{\alpha}.  
 \label{195}
\end{equation}

 The first two components of the  metric and its inverse
have the form
\begin{eqnarray}
   g_{uu}=Ke^{2A(x)}; & g_{uv}=g_{vu}=-\frac{1}{2}e^{2A};&g_{vv}=
0;
 \nonumber \\
  g^{uu}=0;& g^{uv}=g^{vu}=-2e^{-2A};& 
g^{vv}=-4Ke^{-2K} \label{196}
\end{eqnarray}

Non-vanishing components of the Cristoffel symbols are:
 
 \begin{equation}
  \Gamma^{u}_{\alpha u}=\partial _{\alpha} A ,~~
  \Gamma^{v}_{\alpha v}=\partial _{\alpha} A,~~
   \Gamma^{v}_{\alpha u}=\partial _{\alpha} K,
                              \label{197}
 \end{equation}

\begin{equation}
  \Gamma^{\alpha u}_{uu}=-\frac{1}{2}e^{2(A-B)}
  (\partial _{\alpha}K+2K \partial _{\alpha} A ),~~
  \Gamma^{\alpha }_{uv}=\frac{1}{2}e^{2(A-B)} \partial _{\alpha} A ,
                              \label{198}
 \end{equation}

 \begin{equation}
  \Gamma^{\alpha}_{mn}=-\delta_{m}^{n}e^{2(F-B)}\partial_{\alpha} F,
  \label{199}
 \end{equation}

 \begin{equation}
  \Gamma^{m}_{n\alpha}=\delta^{m}_{n}\partial_{\alpha} F,
  \label{1100}
 \end{equation}

\begin{equation}
 \Gamma^{\alpha}_{\beta \gamma}=
 \delta^{\alpha}_{\beta}\partial_{\gamma }B+
 \delta^{\alpha}_{\gamma}\partial_{\beta} B-
 \delta_{\beta}^{\gamma}\partial_{\alpha} B.
 \label{1101}
\end{equation} 

We see that only the Cristoffel symbols containing index $u$  
have changed from those in Appendix A. 
It  is easy to check that only one component of the Ricci tensor 
has  changed

$$ R^{(K)}_{uu}=e^{2(A-B)}\{ -\frac{1}{2}\Delta K 
+K[-\Delta  A +\Delta B +(\partial A)^{2}+
\frac{r(r+1)}{2}(\partial F)^{2}
$$
 \begin{equation}
+3(\partial B)^{2}+2(\partial A\partial B)+
r(\partial A\partial F )+
2r(\partial F\partial B)] +\partial K(\partial A +\partial B
- \frac{r}{2}\partial F) \}
  \label{1102}
 \end{equation}
 The other components do not change
\begin{equation}
 R^{(K)}_{uv}=\frac{1}{2}R^{(K)}_{00}=\frac{1}{2}R_{00},
 \label{1103}
\end{equation}
\begin{equation}
 R^{(K)}_{mn}=R_{mn},~~ R^{(K)}_{\alpha\beta}=R_{\alpha\beta},
 \label{1104}
\end{equation}
as well as the scalar curvature does not change
\begin{equation}
 R^{(K)}=R.
 \label{1105}
\end{equation}
In particular, the LHS of the Einstein equations for 
$uv$ component  
 does not change and we have
 
$$R_{uv}-\frac{1}{2}g_{uv}R=-\frac{1}{2}e^{2(A-B)}[\Delta A+
3\Delta B + r\Delta F  
$$
\begin{equation}
+(\partial A)^{2}+
\frac{r(r+1)}{2}(\partial F)^{2}+3(\partial B)^{2}+2\partial A\partial B+
r\partial A\partial F +
2r\partial F\partial B]
 \label{a311}
\end{equation}
Here we use that 
$$R_{uv}=\frac{1}{4}( R_{00} -R_{11}) =\frac{1}{2} R_{00}. $$

\end{document}